Title: **Bind Recovery of Sparse Factor Structures by Signal Cancellation**

Author: André Achim

ORCID : 0000-0003-2159-6010

Institution : Université du Québec à Montréal

Email : achim.andre@uqam.ca

**Acknowledgements**

The author declares no conflict of interest but wishes to express gratitude for a scholarship from Epilepsy-Canada/Parke-Davis, back in 1988, that allowed pursuing his career in the development of innovative quantitative methods.

**Data availability statement**

Besides a description of SCRoF's use with the illustrative synthetic data, the supplementary material includes these synthetic data, as a (2000,18) matrix without headers, in a file named N2000.csv, as well as five (1000,18) datasets used to observe SCRoF behavior with smaller sample sizes, named N1000_1.csv to N1000_5.csv. The WISC covariance matrix is available in Tabachnick & Fidell's textbook.

**Bind Recovery of Sparse Factor Structures by Signal Cancellation**

Abstract:

Blind factor recovery follows from the principle that the signal of variables exclusive to a factor can be combined in a contrast (weighted sum) that cancels their factor contributions, leaving only a compound of the variables' unique variances. Successful contrasts, uncorrelated with any remaining variable, become the signature of factors with at least two unique indicator variables. Pairwise signal cancellation, usually incomplete for variables affected by different factors, nevertheless succeeds for variables with proportional loadings on two factors, which places three cancelling clusters in the plane of two factors. This is recognized by successful cancellation among variable triplets representing the three clusters. The Signal Cancellation Recovery of Factors (SCRoF) algorithm implements these principles, only requiring that each factor has at least two unique indicators, not even requiring having pre-estimated the number of factors. Alternate sparse factor solutions are obtained through a two significance-threshold strategy. The individually estimated factor loadings and factor correlations of each potential solution are globally optimized for maximum likelihood, yielding a $\chi^2$ indication of compatibility with observed data. SCRoF is illustrated with synthetic data from a complex six-factor structure. Actual data then document that SCRoF can even benefit confirmatory factor analysis when the initial model appears inadequate.

## Introduction

Exploratory factor analysis (EFA, e.g., Thurstone, 1947; Tucker & MacCallum ,1997; Mulaik, 2009; Achim, 2020) applies a common factor model to account for observed correlations among variables in terms of a few common factors that inform these variables. Its main interest is that it provides suggestions about the relationship of the variables to their common factors without requiring a previous hypothesis of such relationship, except for the number of factors. This is especially useful in developing domains of knowledge. Yet, simple solutions are often hard to achieve as EFA allows all factors to inform all variables, leaving it to the user to apply a subsequent rotation to derive a subjectively acceptable and hopefully parsimonious solution (e.g., Howard, 2016).

Given the estimated number of factors, EFA typically proceeds with estimating each variable's unique variance, which is then excluded from the correlation or covariance matrix. The resulting reduced matrix is then decomposed into the preset number of orthogonal dimensions that hopefully delimit the complete and sufficient common factor space. The user must then apply a suitable rotation scheme to produce a sound interpretation of the relationship between factors and variables, sometimes modifying the number of factors to achieve this. Different users may reach different solutions for the same data due to different decisions on (a) the number of common factors to retain, (b) the algorithm to estimate the unique variances subtracted from the data matrix, (c) rotating the principal components of the reduced matrix to meaningful latent variables, or (d) nullifying low factor loadings or low factor correlations.

In principle, sparse matrix solutions (i.e. pattern matrices with many null loadings) should be easier to interpret. A few penalization-based procedures to obtain sparse matrix EFA solutions have been suggested (e.g., Trendafilov et al., 2017). They however all include at least one parameter to be tuned to the data at hand, which carries an extra level of subjectivity. Yang et al. (2024) proposed a different approach to sparse matrix solutions based on the assumption that each variable is exclusive to one of the factors, which does not generally apply.

Confirmatory factor analysis (CFA; Jöreskog, 1967, 1978) yields sparse matrix solutions under the common factor model, but the structure of this model must be pre-specified. Its purpose is to validate whether the data matrix could emanate from the specified structure with its parameters optimized. The optimized model becomes the null hypothesis for the observed data, which is assessed by a $\chi^2$ fit value that should be not significant for the model to be accepted as consistent with the data, given that the sample size was large enough for an incorrect model to yield a significant $\chi^2$. When the initial model is rejected, data-guided exploratory modifications may try to bring the $\chi^2$ fit index to non-significance. Alternately, especially for complex domains where not all relevant sources of information are already known, an approximate model may be accepted, despite the $\chi^2$ fit suggesting it's rejection, when alternate fit indicators less affected by sample size have acceptable values (see Marsh et al., 1988, for independence of fit indices from sample size).

In brief, sparse factor structures are desirable when possible. CFA often tests sparse models but requires pre-specifying the factor structure. It also benefits from a $\chi^2$ fit index useful to judge of the consistency of the optimized model with the observed data. EFA only requires specifying the number of factors but typically returns ambiguous dense solution matrices.

Signal[1] Cancellation Recovery of Factors (SCRoF) is proposed herein as a new approach to the common factor model, hybrid between EFA and CFA, to recover the common factors. SCRoF aims at deciphering the actual factor structure behind the observed correlated variables. Like EFA, it does not require previous knowledge of the solution but, unlike EFA, it does not even require prespecifying the number of factors nor does it proceed through matrix decomposition or need any final heuristic rotation. Its only specific requirement to recover the underlying factors and their correlations is that each factor is expressed in at least two exclusive variables. The approach consists in explaining all variables by cancelling their signal, i.e., the information they received from the common factors, irrespective of the unique variance of each variable.

SCRoF was inspired by graph-network approaches to EFA (Cox & Wermuth, 1993, Golino & Epskamp, 2017) that do not proceed by factorization of a reduced correlation matrix. These approaches conceptualize the correlations as links between pairs of nodes (i.e., variables). Their various algorithms form communities of variables sharing high within-community partial correlations while having as few as possible strong between-community partial correlations. Partial correlation is therefore at the heart of these approaches. Reflection on partial correlation was crucial in the development of SCRoF.

In general, the partial correlation of two variables, *A* and *B*, with respect to a third variable *C* is meant to exclude from both *A* and *B* their common variance also shared with

[1] Here, 'signal' and 'noise' respectively refer to information from shared and unique sources of variances.

*C*, leaving mostly preexisting common variance unique to the pair, along with residual noise. There are two caveats to this. Unless *C* perfectly reflects the underlying common factor, subtracting from both *A* and *B* least-squares determined multiples of *C* only incompletely cancels their information shared with *C*. Furthermore, “decorrelating” both *A* and *B* from *C* injects into each some of the noise part (unique variance) of *C*, inducing some correlation between *A* and *B* due to such shared noise. While the regression residuals of *A* and *B* (i.e., *A* and *B* “decorrelated” from *C*) become orthogonal to variable *C* (its signal plus its noise), these residuals remain correlated even when the three variables only share information from a single common factor.

An extreme case is illustrated in Table 1 with three normalized variables each loading √.5 on their only common factor, resulting in correlations of .5 between *A, B* and *C*. It can be observed that both residuals (decorrelated variables *A*-.5*C* and *B*-.5*C*) have a null correlation with *C* (sum of cross products over the four dimensions) and yet remain correlated .33 due to a coordinate product of .125 on the common factor and another .125 on *C* uniqueness, with the sum of products adjusted for the residuals sums of squares of .75.

Table 1. Coordinates of three variables and their combinations on their common factor and three unique sources of variance demonstrating inadequacy of partial correlations, here retaining a correlation of .33 = (.125+.125)/.75.

| Variable | Orthogonal sources of information | | | | Sum of squares |
|---|---|---|---|---|---|
| | Common | A uniqueness | B uniqueness | C uniqueness | |
| A | √.5 | √.5 | 0 | 0 | 1.0 |
| B | √.5 | 0 | √.5 | 0 | 1.0 |
| C | √.5 | 0 | 0 | √.5 | 1.0 |
| .5C | .5√.5 | 0 | 0 | .5√.5 | .25 |
| A-.5C | .5√.5 | √.5 | 0 | -.5√.5 | .75 |
| B-.5C | .5√.5 | 0 | √.5 | -.5√.5 | .75 |
| (A-.5C)(B-.5C) | .125 | 0 | 0 | .125 | |

This raises the question of how information from a common factor could be correctly removed from a pair of variables to assess whether they share some other common source of information. This is worth a closer diagnostic. In general, decorrelating *A* from *C* amounts to projecting *A* on *C*, keeping only the residual part, which is orthogonal to *C* (as a whole) but not to its signal part. When both variables are normalized to unit length, the projection of *A* on *C* amounts to $r_{AC}\,C$, where $r_{AC}$, the correlation[2] of *A* and *C*, is the product of their respective loadings, namely *a* and *c*, on their common dimension.

[2] Spearman (1904) discussed that the correlation of two observed variables systematically underestimates the correlation between the "true objective value" ('signal' in this manuscript) of the variables each measured with error. He proposed two ways to correct the attenuation, both based on the correlations of independent replications of each variable. He further illustrated that each approach brings the corrected correlation of the "true objective value" of the two variables to 1.0, which is to be expected in absence of noise (although not when one variable reflects a second reliable source of variance besides the common one).

Properly removing this shared signal from *A* would rather require subtracting *a C*/*c* from *A*, where *C*/*c* brings the shared signal in *C* to unit length and multiplication by *a* brings it back to the corresponding signal length in *A*. In brief, partial correlation weight *C* by $r_{AC} = ac$ (both less than 1.0) while the required weight is larger at *a/c*. Thus, what is subtracted from *A* in "decorrelating" it from *C* does not remove all of *A*'s loading on the common dimension, but only a fraction that is $c^2$ too small (although the fraction *c* is unknown), thus incompletely removing the part of *A* on the shared dimension (its signal).

The correct weight, *a/c*, cannot be calculated directly but its correct value resulting in a noise-only contrast can be recognized by absence of correlation of the contrast with any other variable. As for not injecting predictor *C*'s noise into both decorrelated *A* and *B*, this may be achieved by decorrelating the other variable using a different predictor, *D*, also exclusive to the same factor, after having verified that *C* and *D* can form a contrast uncorrelated to any other variable.

These observations led to identifying signal cancellation as a signature for any factor that has at least two exclusive variables. This allows a radically new approach to EFA, here named SCRoF, that does not even require pre-specifying the number of factors. The principle is that two variables exclusive to a factor can always be combined to cancel their signal, retaining only a combination of their noise parts that does not correlate with any other variable.

***Signal Cancellation Recovery of Factors: Principle & Procedure Outline.***

***Signal Cancellation.*** From the geometric point of view on the common factor model, the signal provided by k factors occupies only k dimensions. In particular, the signal of all variables exclusive to the same factor lies a single direction, that of the factor within the multidimensional data space. Signal cancellation thus identifies a factor when a suitable combination of two of its exclusive variables leaves only a combination of their respective unique variances, uncorrelated with any remaining variable. Importantly, two variables along different directions in the common factor space cannot be combined to completely cancel their common signal. Even their best contrast retains correlations with some remaining variables. The complete signal of a variable simultaneously informed by *k* factors can, however, be cancelled by *k* other variables each informed exclusively by one of the *k* factors.

As an example of pairwise signal cancellation, let us say that *A* and *B* are exclusive to the same factor and that *A* loads 0.6 and *B* 0.8 on this factor. These two normalized variables have loadings of 0.8 and 0.6 on their respective sources of unique variance. The weighted difference *A* - (6/8)*B* (called a *contrast*) then cancels the factor contribution to both *A* and *B*. This contrast loads 0.8 on *A*'s uniqueness and -0.45 on *B*' uniqueness. This combination lying outside the common factor space is expected uncorrelated with any variable other than *A* and *B*, with which the contrast shares common noise.

Two variables found to cancel their mutual signal are thus likely exclusively informed by a common factor, but this conclusion is not warranted. Two variables that would load proportionally on the same two factors are also colinear with the origin and can thus mutually cancel their respective composite signals. This direction and those of the two bona fide factors constitute a single plane in the factor space. Yet only two of the three directions in a common plane can be independent factors. One of these directions must be acknowledged as consisting of bifactorial variables.

***Orphan variable exclusion.*** Since signal cancellation is recognized by the absence of significant correlations with any variable outside those in the contrast, it is important procedurally to have first excluded any variable that shares no common variance with the others. Such variables are here called orphan variables. Apparent success at pairwise cancellation would easily result from giving an orphan variable a huge weight relative to that of the other variable. Such contrast would indeed inherit all the non-significant correlations of the orphan variable, but this would incorrectly identify any common factor. SCRoF thus first excludes any variable having no correlation significant at $p \leq .001$ (after correcting for the number of tests). Such variables, if any, will be given null weights in the factor solutions reported.

***Signal cancellation assessment.*** SCRoF uses non-linear optimization to attempt signal cancellation on all pairs of non-orphan variables through contrasts of the form $wA - B$, finding the weight $w$ that minimizes the correlations of the contrast with all remaining variables. The actual minimization bears on the largest absolute value of these correlations. Following optimization, all correlations are combined into a $\chi^2$ value that provides a test statistic for the null hypothesis that the two variables are colinear with the origin and a between-variable distance for subsequent clustering of the variables into common factors.

The reasoning for the $\chi^2$ value is as follows. As a rule, expected null values divided by their standard error become $z$ scores. For expected null correlations, the standard error is $1/\sqrt{(N-1)}$, yielding $z = r\sqrt{(N-1)}$. Also, a squared $z$ score, notably here $z^2 = (N-1)\,r^2$, is a $\chi^2(1)$ (i.e., chi-square with one degree of freedom). The sum of $k$ independent $\chi^2(1)$ is distributed as $\chi^2(k)$ under the null hypothesis. Here $k$ would be the number of correlations of the optimized signal cancellation contrast with the remaining variables, i.e., $k = 8$ for pairwise signal cancellation attempts within a dataset of 10 variables.

***Variable clustering.*** All variables exclusive to the same common factors may be regrouped by cluster analysis. The requirement for clustering of variables into a common putative factor is that all pairs of variables within a cluster should be able to mutually cancel their respective signal. This calls for *complete* clustering, for which the distance between two sub-clusters is the maximal distance between pairs of variables from each cluster. The fusion of two sub-clusters is therefore forbidden by any significant $\chi^2$ between variables from different sub-clusters after adjusting for the number of tests. Indeed, a failure of pairwise signal cancellation, documented by a significant $\chi^2$, implies that at least two common factors are involved in that pair, which forbids the fusion of their respective clusters as a set of variables all exclusively influenced by the same factor.

***Coplanar clusters.*** As already mentioned, successful variable clustering does not guarantee that they share signal from the same common factor. A cluster of two variables having proportional loadings on two factors must however be coplanar with the two clusters of variables exclusive to one or the other of these two bona fide factors. This implies that the number of factors might be less than the total number of clusters emanating from the hierarchical clustering based on pairwise signal cancellation.

To identify the presence of coplanar clusters, all triplets of clusters, at least one of which has only two variables, are tested for possible coplanarity, which involves that any pair of variables from two of these clusters can cancel the signal of any variable from the remaining cluster. Failure of cancellation implies that the three clusters tested occupy three dimensions and thus form three distinct factors. When collinearity is detected between three clusters, the variables of one two-variable cluster are deemed multifactorial, just as are any other variable that entered no cluster. If two or all three coplanar clusters have only two variables, parallel scenarios are developed in which a different two-variable coplanar cluster loses its factor status.

***Multifactorial variables.*** The variables not already associated with a factor are then individually explained through cancellation of their signal by two or more variables that represent distinct factors, again asserting signal cancellation success by lack of correlation of the optimized contrast with the remaining variables. For instance, for a variable *V* whose signal is a sum of two factors respectively represented by variables *A* and *B* and by variables *E*, *F* and *G*, cancellation of the signal of *V* would be achieved by a contrast opposing *V* to a suitably weighted sum of *A* and *E* or of any other pair representing each factor.

Whether the signal cancelling variables (or their respective factors) are correlated or not has no effect on this procedure. Geometrically, signal cancellation amounts to reaching the signal vector of *V* by combining suitable lengths along the two factors that inform *V*, thus using their directions, and then subtracting *V* from the weighted sum of the variables whose respective signal has the direction of their own factor.

***Factor loadings.*** All factor loadings are derived from the optimal signal cancellation weights along with the observed correlations between the variables. For signal cancellation within a pair of normalized variables *A* and *B*, let their respective loadings be *a* and *b*. The expected $r_{AB}$ correlation is then the product of *a* and *b*. That weight *w* cancels the signal in the combination $wA - B$ implies that $wa = b$. Substituting $wa$ for *b* in $r_{AB} = ab$, one gets $r_{AB} = \mathrm{w}a^2$, $a = \sqrt{(r_{AB}/w)}$, and $b = r_{AB}/a$.

For a multifactorial variable $V$ whose signal is cancelled by variables $B$ and $G$, already established to load $b$ and $g$ on their respective factors, the successful contrast is $w_b B + w_g G - V$, such that the loadings are obtained by $w_b b$ and $w_g g$. This is because the optimal weights act by scaling the signal parts of the two cancelling variables.

The average is used when the same variable loading is multiply estimated from its signal cancellation with all possible alternate variables. At this point, the sparse pattern matrix is completed with good estimates of the respective loadings.

***Factor correlations.*** Elaboration of the factor correlation matrix also relies on the SCRoF requirement that each factor is represented by at least two exclusive variables. Although correlation significance for a given pair of factors is based on the first significance test of the canonical correlation applied to the clustered variables of these two factors, the corresponding first canonical correlation underestimates the correlation between these factors due to the presence of residual noise. Rather, following the principle that the correlation between two variables is the product of their respective loading further multiplied by the correlation between their factors, a good correlation estimate is obtained from the weight that brings the vector of products of between-factor loadings closest to the vector of between-factor variable correlations: Placing in vector $O$ the *observed* cross correlations of the variables representing the two factors and in vector $P$ the corresponding *products* of loadings, the factor correlation $r$ minimizes the sum of squares of $O - rP$.

Non-significant correlations are then nullified using a false discovery rate strategy (FDR; Benjamini & Hochberg, 1995). FDR implies ordering the $p$ values for all pairs of factors from the most to the least significant and correcting each $p$ value for assessing the maximum correlation among all not yet assessed correlations. For, say, five factors, there would be 10 correlations. The most significant observed probability, $p_{10}$, is transformed into $1-(1-p_{10})^{10}$. The next most significant probability becomes $1-(1-p_9)^9$, and so on down to a not significant adjusted $p$ value. All following tests are automatically also declared not significant.

As discussed below, two significance thresholds are used, bracketing a zone of higher risk of incorrect decision. All clearly non-significant correlations are first nullified while all clearly significant correlations are maintained. All subsets of correlations that have an adjusted $p$ between the two limits, if any, are nullified in alternate factor correlation matrices.

***Global optimization.*** Mimicking CFA, the resulting sparse pattern factor and correlation matrices of each alternate possible solutions are finally globally optimized for maximum likelihood using their current values as initial parameter estimates and yielding a $\chi^2$ fit index to assess compatibility of the observed data with the model. To prevent purely opportunist fixation of the unconstrained loadings of an orthogonal doublet factor, these are set at the square root of the observed correlation, excluded from the global optimization but included in the model $\chi^2$ fit.

To prevent aberrant solutions, the criterion evaluation function returns a huge value for a parameter combination that brings a variable community above .99 or a factor correlation above .95 or that results in a population correlation matrix with a negative determinant. A solution with a parameter just below these limits is likely incorrect, which is acknowledged by forcing the corresponding $\chi^2$ fit to indicate incompatibility between that solution and the data.

***Alternate Scenarios.*** Alternate factor structure scenarios have already been mentioned about which coplanar clusters to disqualify as a bona fide factor. Alternate scenarios were also implied in nullifying subsets of factor correlations with ambiguous significance. The two-threshold strategy also creates alternative factor structure solutions at the variable grouping stage that specifies the factor pattern matrix, i.e. for deciding on variable clustering and on coplanarity detection, as well as on the number of factors informing multifactorial variables.

***Other statistical considerations.*** To resist both type I (false positive) and type II (false negative) errors, SCRoF uses two statistical thresholds, namely .001 and .25[3]. Obtaining $p \leq .001$ for a statistical test causes rejection of the current null hypothesis. Similarly, $p \geq .25$ makes the scenario consider that the null hypothesis holds. When $.001 < p < .25$, both decisions are considered in alternate processing scenarios. For instance, two sub-clusters with their worst signal cancellation pair yielding $p = .09$ will be grouped together in one scenario and kept as distinct clusters in another.

When several cluster fusions are thus statistically ambiguous, all subsets of ambiguous statistical decisions are considered in separate scenarios. This could however substantially reduce the number of factors when several ambiguous groupings are simultaneously excluded. To prevent exploring solutions with clearly too few factors, a scenario creation is rejected if it would involve fewer than the minimum number of required dimensions as assessed with the Next Eigenvalue Sufficiency Test (NEST; Achim, 2017). This statistically based method was documented not to exceed its nominal type I error rate. The latter is here set at .001, meaning that a *k*-factor model is rejected as insufficient only if the empirical probability of the next eigenvalue (i.e., at rank *k*+1) is less than this low threshold, which strongly limits the risk of overestimating the minimum number of required dimensions.

[3] These default values may be overridden by specifying a pair of thresholds as the final input parameter to SCRoF, e.g. to prevent rejecting variables as orphan when sample size is small.

Many statistical tests in SCRoF apply to an observed maximum. For instance, in deciding if a sub-cluster of two variables may be aggregated with another of three variables, it will require that all six signal cancellation $\chi^2$ involving one variable from each sub-cluster be not significant. The test is thus applied on the most significant of these six $\chi^2$. The associated probability $p$ is then converted to a *net probability* as $1-(1-p)^s$, where $s$, here six, is the number of statistics over which the maximum was retained. With ten variables, for instance, the contrast of each pairwise signal cancelling attempt is correlated with the eight remaining variables; these correlations are transformed into $z^2$ which are summed to form a $\chi^2$ with eight degrees of freedom. If the maximum of the six $\chi^2(8)$ was, say, 27.0 which has an associated $p$ of .0007, this raw probability would become a net probability of $1-.9993^6 = .0042$ that all six $\chi^2$ are at or below the observed maximal value. In such a case, this net probability would cause the aggregation of the two clusters to be both allowed and prevented in parallel scenarios.

## Illustrations

SCRoF performance is first illustrated with simulated data using a voluntary complex factor structure. This will be followed by comparing SCRoF to CFA on data used to illustrate SEM in a popular multivariate statistics manual (Tabachnick & Fidell, 2019).

Before that, however, it is worth noting that the signal cancellation approach is so different from the matrix decomposition approach that it is not constrained by Ledermann's (1937) inequality (from Mulaik, 2009) but only by the requirement of two exclusive variable for each factor. According to this inequality the minimum number of variables necessary to determine two factors should be five. SCRoF, however, will resolve two correlated doublet factors. This is seen, for instance, with all loadings around .8, the factor correlation .5, and N=50. The first illustrative example illustrates, among other SCRoF features, that doublet factors are not problematic provided they correlate with other factors.

**Illustrative Example: Difficult Factor Structure**.

***Condition***. SCFA is first illustrated with an arbitrarily defined 6-factor structure deliberately challenging for standard EFA techniques. It contains two orphan variables, two doublet factors (one correlated and the other orthogonal to remaining factors), a pair of variables with proportional loadings on two factors and another variable loading on three factors. In specifying this factor structure, two concessions were made, namely that each factor should have at least two exclusive indicators, as required for SCRoF, and that each variable should have a community of at least 0.25. The arbitrary factor loadings are given in the left part of Table 1 and their arbitrary specified correlations in the left part of Table 2. The right part of these Tables gives the apparently preferable solution (see discussion). The side-by-side presentations are aimed at facilitating appreciation of SCRoF performance.

Table 2. Population (left) and SCRoF 'preferred' solution (right) for synthetic data with N=2000.

| Variable rank | Population common factors | | | | | |
|---|---|---|---|---|---|---|
| | 1 | 2 | 3 | 4 | 5 | 6 |
| 1 | **.5** | 0 | 0 | 0 | 0 | 0 |
| 2 | **.6** | 0 | 0 | 0 | 0 | 0 |
| 3 | **.55** | 0 | 0 | 0 | 0 | 0 |
| 4 | 0 | **.5** | 0 | 0 | 0 | 0 |
| 5 | 0 | **.6** | 0 | 0 | 0 | 0 |
| 6 | 0 | **.5** | **.75** | 0 | 0 | 0 |
| 7 | 0 | **-.4** | **-.6** | 0 | 0 | 0 |
| 8 | 0 | 0 | **.5** | 0 | 0 | 0 |
| 9 | 0 | 0 | **.6** | 0 | 0 | 0 |
| 10 | 0 | 0 | 0 | **.5** | 0 | 0 |
| 11 | 0 | 0 | 0 | **.6** | 0 | 0 |
| 12 | 0 | **.4** | **.6** | **.4** | 0 | 0 |
| 13 | 0 | 0 | 0 | 0 | **.5** | 0 |
| 14 | 0 | 0 | 0 | 0 | **.8** | 0 |
| 15 | 0 | 0 | 0 | 0 | 0 | **.5** |
| 16 | 0 | 0 | 0 | 0 | 0 | **.8** |
| 17 | 0 | 0 | 0 | 0 | 0 | 0 |
| 18 | 0 | 0 | 0 | 0 | 0 | 0 |

| Variable rank | SCRoF solution common factors | | | | | |
|---|---|---|---|---|---|---|
| | 1 | 2 | 3 | 4 | 5 | 6 |
| 1 | **.54** | 0 | 0 | 0 | .0 | .0 |
| 2 | **.58** | 0 | 0 | 0 | 0 | 0 |
| 3 | **.54** | 0 | 0 | 0 | 0 | 0 |
| 4 | 0 | **.47** | 0 | 0 | 0 | 0 |
| 5 | 0 | **.58** | 0 | 0 | 0 | 0 |
| 6 | 0 | **.47** | **.76** | 0 | 0 | 0 |
| 7 | 0 | **-.36** | **-.63** | 0 | 0 | 0 |
| 8 | 0 | 0 | **.53** | 0 | 0 | 0 |
| 9 | 0 | 0 | **.61** | 0 | 0 | 0 |
| 10 | 0 | 0 | 0 | **.45** | 0 | 0 |
| 11 | 0 | 0 | 0 | **.59** | 0 | 0 |
| 12 | 0 | **.39** | **.60** | **.42** | 0 | 0 |
| 13 | 0 | 0 | 0 | 0 | **.45** | 0 |
| 14 | 0 | 0 | 0 | 0 | **.81** | 0 |
| 15 | 0 | 0 | 0 | 0 | 0 | **.61** |
| 16 | 0 | 0 | 0 | 0 | 0 | **.61** |
| 17 | 0 | 0 | 0 | 0 | 0 | 0 |
| 18 | 0 | 0 | 0 | 0 | 0 | 0 |

Table 3. Arbitrary correlations applied to the factors in the population (left) and in the SCRoF 'preferred' solution (right).

| | | | | | |
|---|---|---|---|---|---|
| 1 | **.4** | **-.3** | 0 | **.4** | 0 |
| .4 | 1 | 0 | **-.3** | **.3** | 0 |
| -.3 | 0 | 1 | 0 | **-.5** | 0 |
| 0 | -.3 | 0 | 1 | 0 | 0 |
| .4 | .3 | -.5 | 0 | 1 | 0 |
| 0 | 0 | 0 | 0 | 0 | 1 |
| 1 | **.48** | **-.26** | 0 | **.35** | 0 |
| .48 | 1 | 0 | **-.29** | **.29** | 0 |
| -.26 | 0 | 1 | 0 | **-.50** | 0 |
| 0 | -.29 | 0 | 1 | 0 | 0 |
| .35 | .29 | -.50 | 0 | 1 | 0 |
| 0 | 0 | 0 | 0 | 0 | 1 |

The population signal eigenvalues (exclusive of unique variances) are 2.51, 1.50, 0.89, 0.77, 0.52, and 0.26. The eigenvalues of the population correlation matrix are 2.96, 2.13,

1.42, 1.40, 1.13, 1.00, 1.00, 0.93, 0.73, 0.72, 0.70, 0.70, 0.67, 0.60, 0.59, 0.58, 0.47, and 0.29. The two 1.00 for the sixth and seventh population eigenvalues are due to the orphan variables. Given the complexity of this factor structure, sample size was set at N=2000 to clearly illustrate SCRoF.

Subsequent analyses of five datasets with N=1000 (also available as supplementary material) provided essentially the same results, with the correct solution having $p > .39$ in four samples. The remaining dataset contained the correct solution with $\chi^2(94) = 119.21$, $p = .041$, which happened to be the best fit among all reported scenarios.

***Results.*** Details of the SCRoF output are provided as supplementary material. The initial clustering of pairwise cancellation gave seven clusters (see Figure 1). Clear coplanarity was detected, with no sign of cancellation failure ($p = .85$) of the variables of one coplanar cluster by the other two. The coplanar clusters consisted of (v4, v5), (v6, v7) and (v8, v9). Although scenarios with v4 and v5 made bifactorial were present in the reports of all five datasets with N=1000, the present test data yielded no such scenario statistically consistent with the data, which is likely due to the correlation estimated to .85 between the other two clusters taken as the factors subtending this plane.

Making v8 and v9 bifactorial implied a correlation of .54 between the remaining two coplanar clusters. Rather making v6 and v7 bifactorial implied orthogonality between the remaining two clusters, justifying preference for this factor structure. We therefore considered the cluster of v4 and v5 as population factor F2 and that of v8 and v9 as factor F3.

There were four scenarios with v6 and v7 as bifactorial. Two of them, differing only by one factor correlation, were equivalent to the other two but were reached through different paths, one with v6 and v7 not clustered together and one where their coplanar cluster became bifactorial. Nullifying the ambiguous correlation of -.056 between F1 and F6 gave one extra degree of freedom at the cost of a modest increase of 1.03 in the associated $\chi^2$. The solution preferred for the orthogonality of one of its pairs of factors and for one extra degree of freedom has $\chi^2(94) = 109.20$, $p = .135$. Its clustering dendrogram if depicted in Figure 1. This happens to be the correct solution although we would not know that.

Figure 1. Variable clustering dendrogram based on pairwise signal cancellation, for the preferred solutions in which v6, v7 and v12 are multifactorial. The initial cluster of v6 and v7 is marked by grey lines.

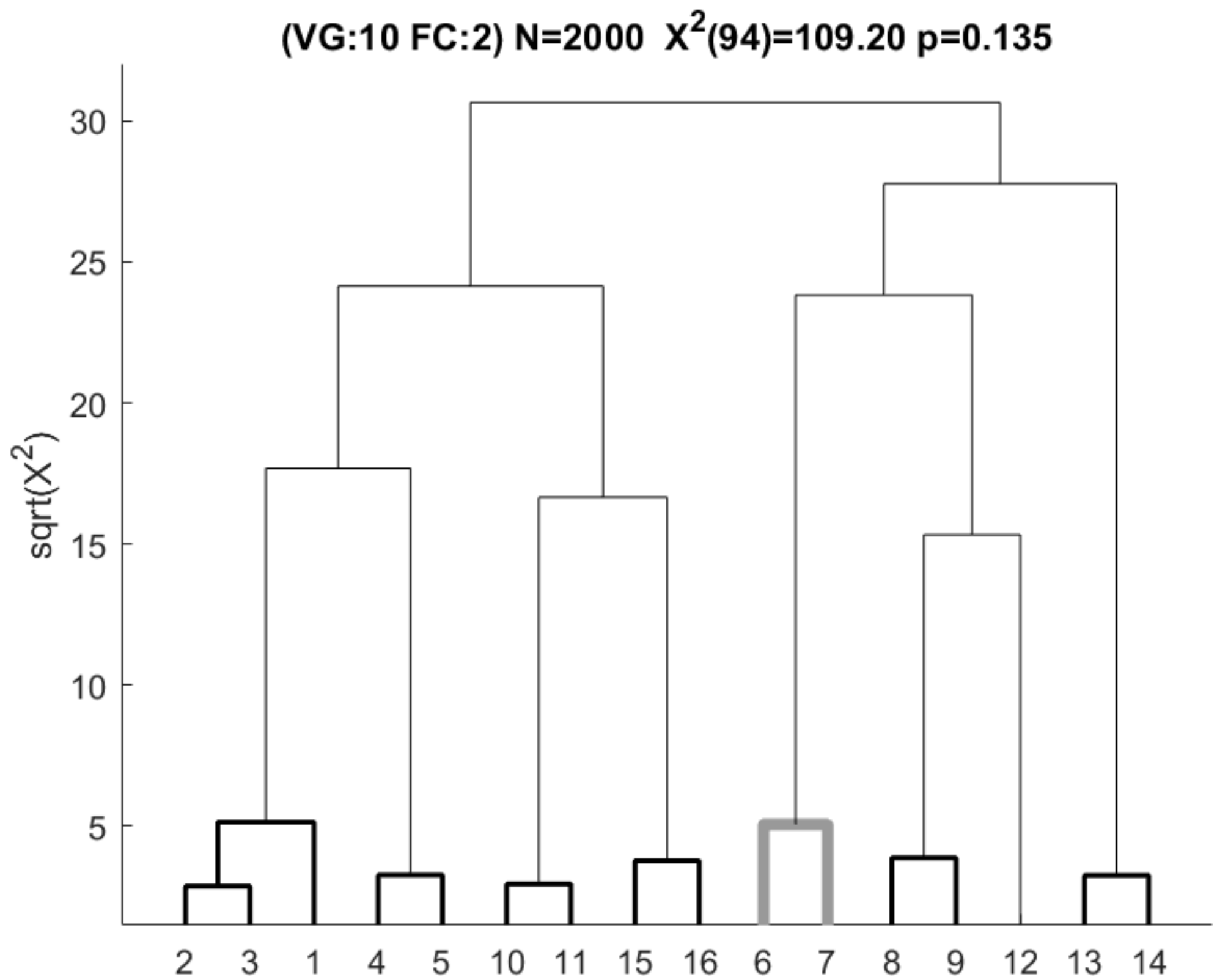

***Conclusion.*** Although SCRoF requires no user preference or guess for analyzing the data, it may provide several solutions statistically consistent with the data. It is then appropriate to discuss these alternatives, providing reasons for preferring one of them. For this complex synthetic example, we retain that the correct solution was among the scenarios qualified at $p > .05$. A statistically viable alternative existed for delimiting the three clusters plane. As the two factors subtending this plane would correlate .54, the two-orthogonal-factor description of the plane was preferred in absence of theoretical considerations that would suggest otherwise.

### *Real Data Example*

We next illustrate (a) that large sample sizes are not a particular requisite of SCRoF compared to alternate EFA methods and (b) that signal cancellation can bring extra constraints on CFA solutions when the original model needs modification.

In Chapter 14 of Tabachnik & Fidell (2019), author J.B. Ullman used 11 WISC subscales data from 177 learning disabled children to illustrate structural equation modelling. Two outlier cases, one univariate and one multivariate, were removed. The data were initially modelled as reflecting two correlated factors, where the first six subscales are taken as pure indicators of Verbal intelligence and the last five as pure indicators of Performance intelligence. This model did not fit the data, with $\chi^2(43) = 70.2$ ($p$ =.0054). LISREL suggested adding a link from Performance intelligence to Comprehension, which brought the fit to $\chi^2(42) = 60.3$ ($p$ =.033). As Coding did not load significantly on Performance intelligence, its further removal yielded the final model with $\chi^2(33) = 45.0$ ($p$ = .08).

***Hypothesis.*** Adopting the signal cancellation point of view, this solution implies that no other variable could singly cancel the bifactorial signal of Comprehension, it being the only bifactorial variable in the model. SCRoF should confirm this prediction if the solution is correct with all other variables indeed unifactorial. The SCRoF analysis was based on the covariance matrix reported in Tabachnick & Fidell (2019), pages 576-577.

Figure 2. Variable clustering dendrogram expressing pairwise signal cancellation among the 10 retained WISC subscales, which clearly shows the mutual signal cancellation of Comprehension (v2) and Similarities (v4).

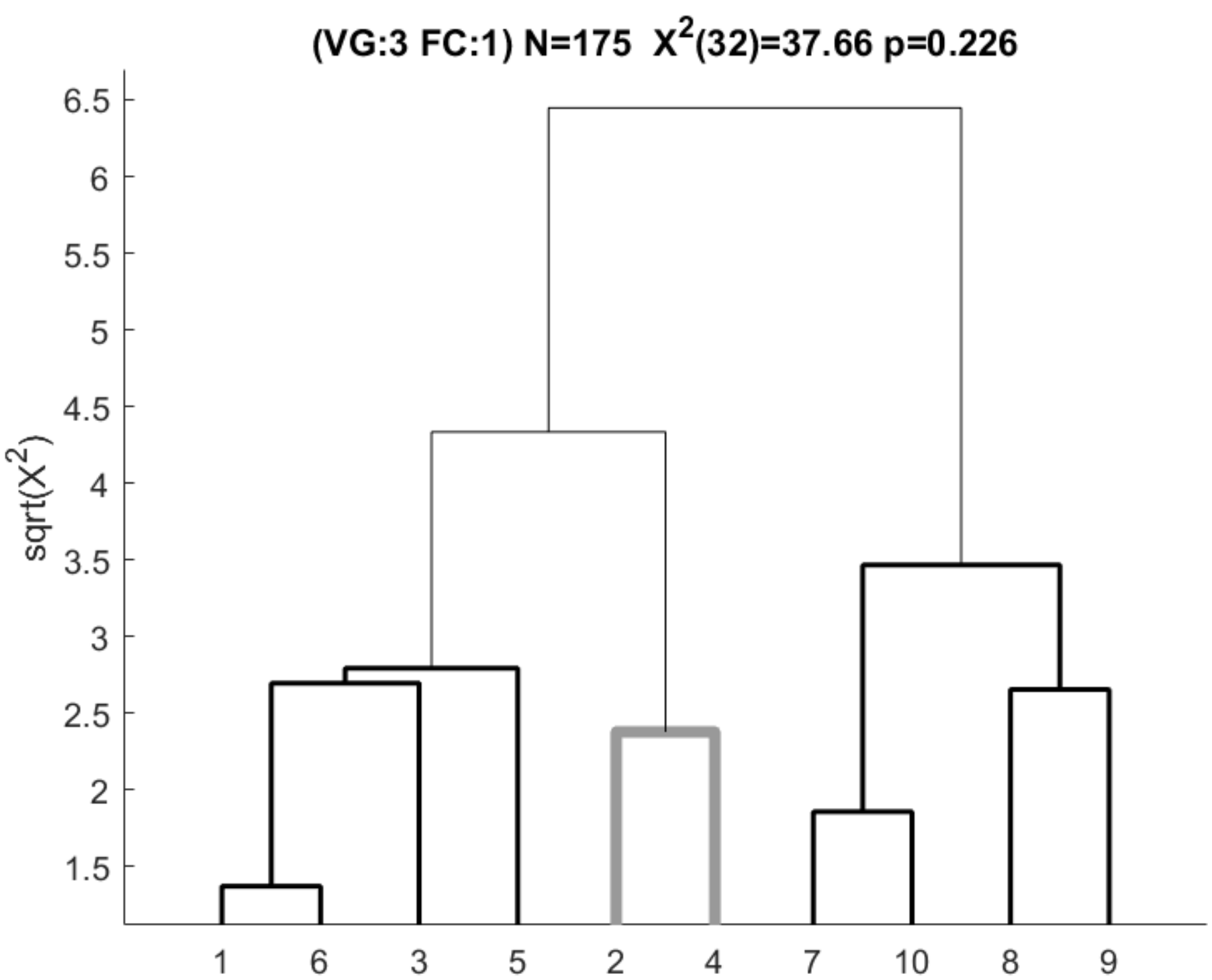

***Results.*** SCRoF excluded Coding as an orphan variable and retained three scenarios. As illustrated in Figure 2, v2 and v4 (respectively Comprehension and Similarities) unambiguously cancel their respective signal in this sample ($\chi^2(8) = 5.65$, $p = .69$). One scenario with $\chi^2(34) = 55.0$, $p = .013$, corresponds to the initial CFA model, exclusive of Coding, with all indicators unique to their factor. The other two scenarios rejected the clustering of the Comprehension and Similarities cluster with that of the other indicators of Verbal intelligence. One scenario, with $\chi^2(32) = 37.7$, $p = .23$, had the Comprehension and Similarities cluster coplanar with the Verbal and Practical intelligence factors, these correlating .48. Compared to the initial model, the two extra degrees of freedom used decreased $\chi^2$ by 17.3, $p = .00017$. This is the preferred solution as the third scenario, despite a similar fit ($\chi^2(32) = 38.3$, $p = .21$), proposed a less parsimonious three factor model in which the extra doublet factor informing Comprehension and Similarities would correlate .91 with the Verbal and .75 with the Performance factors. The preferred two-factor solution is presented in Table 4, along with the final textbook solution.

Table 4. Final textbook CFA solution to the 10 WISC subscales (after deleting Coding) and best SCRoF solution. Thei factor correlation is.59 for CFA and .48 for SCRoF.

| Variable | | CFA | | SCRoF | |
|---|---|---|---|---|---|
| Rank | Name | Verbal | Performance | Verbal | Performance |
| 1 | Information | **.78** | 0 | **.79** | 0 |
| 2 | Comprehension | **.50** | **.30** | **.47** | **.35** |
| 3 | Arithmetic | **.56** | 0 | **.57** | 0 |
| 4 | Similarities | **.70** | 0 | **.55** | **.25** |
| 5 | Vocabulary | **.78** | 0 | **.78** | 0 |
| 6 | Digit Span | **.40** | 0 | **.41** | 0 |
| 7 | Picture Completion | 0 | **.62** | 0 | **.63** |
| 8 | Picture Arrangement | 0 | **.45** | 0 | **.45** |
| 9 | Block Design | 0 | **.67** | 0 | **.64** |
| 10 | Object Assembly | 0 | **.58** | 0 | **.58** |

***Conclusion.*** By finding unambiguous signal cancellation of Comprehension by another variable, namely Similarities, SCRoF invalidates the earlier SEM solution for this sample by indicating that Comprehension and Similarities have proportional enough loadings on both the Verbal and Performance factors for these to mutually cancel their signal. This illustrates that signal cancellation brings a new source of information to factor analysis.

## Discussion

The principle of signal cancellation allows a radically new approach to EFA in which factors are individually identified from pairs of their unique indicator variables. Once the unifactorial variables are associated with their respective common factors through pairwise signal cancellation, the remaining multi-factor dependent variables are explained by cancelling their composite signals by subsets of unifactorial variables. Contrary to the standard EFA model that allows all factors to affect all variables, this approach naturally produces sparse solutions. These rotation-free solutions are blindly produced, without any user intervention not event to specify the number of factors.

The SCRoF two-threshold approach, however, may yield several solutions consistent with the data. Thus, the user's judgement and preferences expurgated from classical EFA procedures may reappear following SCRoF analysis in discussing the relative merit of alternate solutions. This is not unlike standard EFA in exploring alternate rotations. When more than one viable scenario is reported, the recommended practice is to discuss the reason to prefer one over any other. Nothing however prevents simply acknowledging that the data are compatible with a few theoretically distinct solutions, especially if this suggests further research to select among them, like to document the necessity of an extra parameter.

The present SCRoF report already indicates that, along with the $\chi^2$ fit probability of the data given the model, assessing the merits of alternate solutions should consider model parsimony and between-factors correlations. The 6-factor synthetic example illustrated that the correct solution was among those not rejected by the $\chi^2$ test, although other solutions could have presented better fits. For instance, a developing version of SCRoF that still allowed aberrant solutions (by not preventing an orthogonal doublet factor to further inform a multivariate variable nor penalizing solutions with loadings topped at .99) produced a solution whose $\chi^2$ fit exceeded that of the correct one by 16.7 for 2 degrees of freedom ($p < .0003$). In that case, the "statistically better" solution could be rejected as aberrant on the two grounds named above, but this provided a lesson that it could be misleading to blindly prefer a solution simply because its is significantly better than another one also compatible with the data. More experience and debates among experts will be required on the ensuing question of how large a fit improvement must be, between two otherwise sound and statistically acceptable models, to carry its preference over the more parsimonious alternative. In the real data example, the model with two bifactorial variables would not have been so easily preferred if the initial model (all variables depending on a single factor) had shown a fit probability above .05, even if the former appeared significantly better with, say, $p = .03$. Such result would rather call for further research.

The SCRoF prerequisite that each factor has two unique indicator variables can be falsified by failure to explain some variables through signal cancellation. But this prerequisite should not be considered proven by the occurrence of scenarios compatible with the data. In a bifactor model (Holzinger & Swineford, 1935), for instance, each variable is considered affected by two factors, a general factor that is common to all variables and a factor specific to each non-overlapping subset of variables, where all factors are assumed orthogonal. Fairly equal loadings of the general and specific factors would imply that each subset of bifactorial variables is nearly colinear with the factor space origin. SCRoF would then provide a solution that satisfactorily explains the data as depending only on the set of specific factors, although these would be highly correlated. In a sense, the underlying bifactor model is a solution that further explains what causes the factors to be correlated[4]. It remains the responsibility of the user to assume that each factor has at least two unique variables. Should however the loadings of the general and specific factors deviate markedly form proportionality, SCRoF would be expected to fail to produce any solution statistically consistent with the data if sample size is sufficiently large.

The main problem with SCRoF solutions is to recognize incorrect simplified solutions when sample size is not large enough. When all factors are positively correlated and expressed in few indicators with moderate loadings, like the most difficult simulation conditions of Haslbeck & van Bork (2024), SCRoF may even report only one statistically acceptable solution that has fewer that the correct number of factors. It may be wished that further work identifies diagnostic indicators of incorrect solutions. Early exploration in this direction however showed that the sequence of eigenvalues of the data, as well as their pairwise signal cancellation dendrograms patterns, may appear completely unremarkable among those of surrogate data generated with an incorrect factor structure not rejected by SCRoF due to too few cases. Thus, besides the assumption of two exclusive variables for each factor, SCRoF crucially depends, like CFA and EFA, on the assumption of a large enough sample size.

[4] In general, accepting factor correlations without explaining them implies admitting incomplete knowledge of the phenomenon studied.

Besides its two-unique-variable requirement, SCRoF also relies, like other factoring approaches, on the assumption of additive effects of the common factors on multivariate indicators. Assuming that the underlying factor scores are symmetrically distributed, skewed variable distributions should be attributable to the measurement instrument (surface skewness), which could distort factor level additivity. A ghost factor appears, for instance, to account for the higher correlations between same-skewness variables than opposite-skewness variables that all depend on the same factor (e.g. Brandenburg, 2024). It is therefore good practice to apply symmetrizing transformations before factoring the data, be it by EFA, by SCRoF or by CFA. Preliminary inspection of the data and eventually variable transformations remain strongly recommended, even for SCRoF.

No consideration has yet been given to missing data handling. Further work could assess SCRoF robustness to different proportions of missing data and offer proposal for adjusting the sample size associated with the resulting correlation matrix. But the two-threshold approach would likely immunize SCRoF from serious distortions.

Similarly, the assumption of distribution normality behind the global $\chi^2$ optimization does not appear crucial for SCRoF, given its two-threshold approach that shields it from imprecise probability estimates. Again, transforming strongly skewed distribution into more symmetrical ones is recommended, here to alleviate potential $\chi^2$ bias effects, as sums of any quantity converge to normality faster for symmetrical than for skewed distributions.

Finally, it seems that the signal cancellation approach could be extended to provide sparse EFA solutions when some or all factors do not have two exclusive indicator variables. The signal of variables left unexplainable by sets of variables that pairwise cancelled could nevertheless be cancelled by other variables sharing some of their factor signals. In such extension of SCRoF, the common factor space would be reliably delimited although the factors themselves would remain unconstrained within that space.

**Availability**

MATLAB code for SCRoF is available on MATLAB File Exchange at https://www.mathworks.com/matlabcentral/fileexchange/177674-signal-cancellation-recovery-of-factors.

P.-O. Caron, from Université TÉLUQ, (Pier-Olivier.Caron@Teluq.ca) should shortly publish on GitHub (https://github.com/quantmeth/SignalCancellation) and eventually on CRAN an **R package**, called SignalCancellation, that includes SCRoF and a true partial correlation function.

*Supplementary material: Using SCRoF*

The present analysis was performed with the MATLAB version of SCRoF. Its messages are in French, but it should be easy to figure out their meanings. The upcoming R version of SCRoF might set all messages in English.

SCRoF receives data as input. Here, 'dat' was a matrix of size (2000,18). Since SCRoF operates on the data correlation matrix, its input may alternately be a correlation or a covariance matrix along with sample size. In the MATLAB version, the input matrix has no variable-name header; the variables are only referred to by their rank. The SCRoF function returns a structure (here named 'AS') that contains the details of all explored scenarios along with several intermediate results. SCRoF terminates by printing a list of scenarios of interest, along with details that characterize each. These are all the explored scenarios compatible with the data ($p$ >.001) except for the first listed scenario that is reported irrespective of its fit with the data. This first scenario has all clusters from pairwise signal cancelation, even in the presence of clear coplanarity, and is included to ease access to the correlations between all initial clusters. An associated command, described later, prints the solution associated with a given line number. SCRoF printed output for this complex illustrative example is here presented and commented. Simpler data structures, however, usually come with much fewer output lines.

```
>> AS=SCRoF(dat);
NEST indique au moins 5 fct, suggère 5, AP_50_95 suggèrent 5 5 fct
Scénarios d'intérêt:
1:   p=0.0000 X2(97)=1622.830 VG1 FC1  7f Grappes VG2 plan:(0.828:2,3,4)
2:   p=0.1122 X2(93)=109.831 VG2 FC1  6f Initial
3:   p=0.1345 X2(93)=108.169 VG3 FC1  6f Grappes VG1
```

4: p=0.1352 X2(94)=109.203 VG3 FC2 6f "
5: p=0.0612 X2(93)=114.934 VG5 FC1 6f bf:8,9 Coplan VG1
6: p=0.1253 X2(93)=108.827 VG6 FC1 6f bf:6,7 Coplan VG1
7: p=0.1461 X2(92)=106.309 VG8 FC1 6f MultiSatur VG2
8: p=0.0747 X2(94)=114.432 VG9 FC1 6f bf:8,9 Coplan VG4
9: p=0.0751 X2(95)=115.484 VG9 FC2 6f "
10: p=0.1334 X2(93)=108.249 VG10 FC1 6f bf:6,7 Coplan VG4
11: p=0.1352 X2(94)=109.203 VG10 FC2 6f "
12: p=0.0717 X2(92)=112.545 VG12 FC1 6f bf:8,9 MultiSatur VG5
13: p=0.1483 X2(92)=106.170 VG13 FC1 6f bf:6,7 MultiSatur VG6
14: p=0.0714 X2(93)=113.695 VG15 FC1 6f bf:8,9 MultiSatur VG9
15: p=0.0723 X2(94)=114.696 VG15 FC2 6f "

An associated procedure, SCRoFreport, prints the matrices of factor loadings and correlations when provided with the SCRoF output structure and a scenario line number. It also paints the clustering dendrogram of the specified scenario in which variables within a common cluster are linked with wider lines. For a cluster that the scenario made bifactorial, these lines are grey rather than black. A variable that never belonged to a cluster starts with a thin black line.

***Interpreting scenario lines.***

Before discussing the merits of these various possible solutions, it is appropriate to describe the appearance of each line. The first output line appears on the screen early in SCRoF processing. It estimates the number of factors using NEST (Achim, 2017). The minimum number of factors is set by rejecting with $p < .001$the null hypothesis that k factors are sufficient, meaning that the eigenvalue at rank k+1 of the data is larger than all corresponding eigenvalues from 1000 surrogate datasets generated with a suitable k-factor model. SCRoF will explore no scenario that would include fewer factors than this minimum number. Additional estimations on the same line are given for information only. One is the usual, less conservative, NEST suggestion rejecting insufficient models at $p < .05$. This is followed by two parallel analysis (Horn, 1965) suggestions respectively using the $50^{th}$ and $95^{th}$ centiles of 1000 datasets generated with a null factor model. For the present data, all those indices underestimated to 5 the number of factors.

Each "Scénarios d'intérêt" output line starts with the scenario identifier number followed by its fit probability, degrees of freedom and $\chi^2$ values. Then 'VG' is followed by a number that indicates the corresponding entry into the field VG (for Variable Grouping) of output structure AS. This is provided to ease consultation of the AS structure if required. These VG numbers should not be confused with the printed scenario identifiers. This is followed by 'FC' and a factor correlation variant rank. 'FC1' has all correlations significant at $p < .25$. Further 'FC' values for the same VG number involve withdrawal (nullifying) of some subset of correlations whose significance lies between the two statistical thresholds, hence with a correspondingly larger number of degrees of freedom. The line continues with the number of factors in the scenario (e.g., '6f'). If a scenario factor structure implied a variable correlation above .99 in absolute value in the population, its number of factors would be expressed as negative.

Lines with 'FC1', i.e. those introducing a new variable grouping, are completed by further information concerning that variable grouping. For a scenario in which coplanarity is acknowledged, the initially clustered variables that became bifactorial are listed next (e.g. 'bf:8,9'). Then comes an indication of the procedure that created the scenario, followed (except for procedure 'Initial') by the variable grouping from which it was derived. This initial scenario, here listed in line #2, consists exclusively of the unambiguous clusters (i.e., all clustering with $p > .25$).

Line #1, that here ends with '7f Grappes VG2 plan:(0.828:2,3,4)', describes the first scenario produced with the maximum number of clusters observed and with possible coplanarity, if any. This was created for the present data while managing the clustering ('Grappes') parameters, where a cluster was added upon acceptance of the clustering of two variables (later discovered to be v6 and v7) that mutually cancelled their signal ambiguously, i.e., with a probability between the two thresholds. The AS output structure contains information that this was $p = .0294$. The 'VG2' part of this line indicates that this was derived from the scenario that has 'VG2' immediately after its fit values, namely here line #2.

Line #1 provides some extra information about its clusters that share a coplanar relationship. This information starts with the adjusted worse probability of signal cancellation of the variables in the cluster named last by those in the other two clusters. Here, clusters 2, 3 and 4 were unambiguously coplanar with $p = .828$. This being above the .25 threshold, coplanarity is not in doubt, meaning that the seven *clusters* imply fewer than seven *factors*. This makes signal cancellation of the variables not belonging to any of its seven clusters both non-unique and irrelevant. Their loadings being left null yield a huge observed $\chi^2$ value, signalling that this seven-factor scenario is not an option. Note that scenario 'VG4' is not listed. It differs from 'VG1' by further aggregating v1 to the cluster of v2 and v3. Having seven clusters as well and with no attempt to explain v12, it does not qualify as a scenario of interest.

Two other procedures that may create new scenarios are designated by ‘Coplan’ or by ‘MultiSatur’. The former applies to a scenario variant that designate which of three coplanar clusters was considered not a bona fide factor. That there is no reported scenario line containing ‘bf:4,5  Coplan’ does not mean that such scenario were skipped from VG1 and from VG4. They rather both turned out bad fits ($p < .001$) to the data and are just not reported. The ‘MultiSatur’ indication refers to signal cancellation attempts of variables not included in a cluster. This is the case for v12, but also for v1 from VG1 and for v6 and v7 in scenarios in which their clustering was rejected for ambiguous pairwise cancellation with $p < .25$. Cancellation of their signal is first attempted using variables representing a pair of clusters. If this results in the best cancellation associated with a probability between the two thresholds, a new scenario is created for attempted cancellation by three clusters, unless the scenario acknowledged only two factors.

***Selecting a solution.***

There is no necessity to retain a single solution when two of them appear having comparable merit. In absence of theorical plausibility, preference could go to parsimony, including preference for orthogonality over correlation of a pair of factors, or for lower correlations among the retained factors when coplanarity is detected, as is the case here. It is then highly relevant to inspect the correlations among all initial clusters (putative factors). These correlations are reported, along with the irrelevant scenario #1 factor loadings and its irrelevant clustering dendrogram, by calling SCRoFreport with the output structure name and scenario rank 1 as input parameters. In the relevant part of the output, the correlations among the three coplanar clusters, namely 2, 3 and 4, were underlined.

```
>> SCRoFreport(AS,1);
[…]
fCorr:
```

| | 1 | 2 | 3 | 4 | 5 | 6 | 7 |
|---|---|---|---|---|---|---|---|
| 1 | 1.000 | 0.447 | 0.000 | -0.216 | 0.000 | 0.264 | 0.000 |
| 2 | 0.447 | 1.000 | 0.538 | 0.000 | -0.304 | 0.275 | 0.000 |
| 3 | 0.000 | 0.538 | 1.000 | 0.848 | 0.000 | -0.254 | 0.000 |
| 4 | -0.216 | 0.000 | 0.848 | 1.000 | 0.000 | -0.464 | 0.000 |
| 5 | 0.000 | -0.304 | 0.000 | 0.000 | 1.000 | 0.000 | 0.000 |
| 6 | 0.264 | 0.275 | -0.254 | -0.464 | 0.000 | 1.000 | 0.000 |
| 7 | 0.000 | 0.000 | 0.000 | 0.000 | 0.000 | 0.000 | 1.000 |

That the correlation between clusters 2 and 4 is null makes it attractive to consider the plane as consisting of these two orthogonal factors, with v6 and v7, constituting cluster 3, lying in their plane. If clusters 2 and 3 were the factors of this plane, they would correlate .54. Selecting clusters 3 and 4 as factors would have them correlating 0.85. From here on, the factor clusters will be referred to by their corresponding name in the generating model. For instance, we acknowledge that v6 and v7 load on simultaneously F2 and F3.

Scenario #2 has v1, v6, v7 and v12 not part of a cluster. Its factor matrix, obtained from SCRoFreport, has v1 loading .53 on F1 and .007 on F4. Scenario #7, emanated from VG2 with the 'MultiSatur' mention, has v1 loading .52 on F1, -.044 on F3 and -.007 on F4. These two scenarios, #2 and #7, having a hierarchical relationship, their 3.52 difference in $\chi^2$ does not constitute a significant improvement for the extra degree of freedom. Similarly, the differences in $\chi^2$ for removing a borderline correlation between #3 and #4, as well as between #10 and #11, do no justify keeping the correlation. This excludes #3 and #10 as valuable solutions.

Scenario #4 emanated from unreported VG4 that accepted the clustering of v1 with v2 and v3. It thus differs from #2 by having v1 loading exclusively on F1. The $\chi^2$ difference is less than 1.0, which disqualifies scenario #2. Scenario #6 had v1 not clustered with v2 and v3 but v6 and v7 clustered together and later acknowledged as coplanar. Variable v1 received a loading of -.016 on F4. This is very similar to the rejected scenario # 2. Scenario #13 is a variation of #6 with a further loading of -.041 on F3. Compared to the identical scenarios #4 and #11, reached in two different paths, the two extra parameters of solution #13 are not justified by the 3.04 $\chi^2$ difference for two degrees of freedom.

After having discarded the other solutions, the preferred solution has v1 depending exclusively on F1, v6 and v7 bifactorial on F2 and F3, and v12 dependent on F2, F3 and F4.